\documentclass[10pt, twocolumn, aps, prd, amssymb, amsmath, superscriptaddress, eqsecnum, nofootinbib, notitlepage, tightenlines, longbibliography]{revtex4-1}

\usepackage[CJKbookmarks, pdftex, bookmarksnumbered, bookmarksopen, colorlinks, citecolor=crimson, linkcolor=blue]{hyperref}

\usepackage[utf8]{inputenc}
\usepackage[T1]{fontenc}
\usepackage{graphicx}
\usepackage{hyperref}
\usepackage{verbatim}
\usepackage{amssymb}
\usepackage{amsmath}
\usepackage[english]{babel}
\usepackage{amsthm}
\usepackage{mathrsfs}
\usepackage{mathtools}
\usepackage{amsfonts,bbm}
\usepackage{amssymb,amscd}
\usepackage{dsfont}
\usepackage{soul}
\usepackage{enumitem}
\usepackage{tikz,bbm,bm}
\usepackage{cancel}
\usepackage[all,cmtip,color]{xy}
\usepackage{braket}
\usepackage{xypic,color}
\usepackage[many]{tcolorbox}
\usepackage{nccmath} 
\usepackage[normalem]{ulem}
\usepackage{float}
\usepackage[format=plain,font=footnotesize]{caption}
\usepackage{subcaption}

\allowdisplaybreaks

\newcommand{\Ignore}[1]{}

\newcommand{\be}{\begin{equation}}
\newcommand{\ee}{\end{equation}}
\newcommand{\bes}{\begin{eqnarray}}
\newcommand{\ees}{\end{eqnarray}}

\renewcommand\ket[1]{{|{#1}\rangle}}

\definecolor{crimson}{rgb}{0.86, 0.08, 0.24}
\definecolor{persianblue}{rgb}{0.11, 0.22, 0.73}
\definecolor{cobalt}{rgb}{0.0, 0.28, 0.67}
\definecolor{green(pigment)}{rgb}{0.0, 0.65, 0.31}
\definecolor{mediumseagreen}{rgb}{0.2, 0.6, 0.65}
\definecolor{lightgray}{rgb}{0.83, 0.83, 0.83}
\definecolor{anti-flashwhite}{rgb}{0.95, 0.95, 0.96}

\definecolor{crimson}{rgb}{0.7, 0.08, 0.24}

\newcommand{\partdif}[2]{\frac{\partial #1}{\partial #2}}

\usepackage{scalerel,stackengine}
\stackMath
\newcommand\reallywidehat[1]{%
\savestack{\tmpbox}{\stretchto{%
  \scaleto{%
    \scalerel*[\widthof{\ensuremath{#1}}]{\kern-.6pt\bigwedge\kern-.6pt}%
    {\rule[-\textheight/2]{1ex}{\textheight}}%WIDTH-LIMITED BIG WEDGE
  }{\textheight}% 
}{0.5ex}}%
\stackon[1pt]{#1}{\tmpbox}%
}
\def\beqa{\begin{eqnarray}}
\def\eeqa{\end{eqnarray}}
\def\bean{\begin{eqnarray*}}
\def\eean{\end{eqnarray*}}

\newcommand{\dd}{\mathrm{d}}
\usepackage[normalem]{ulem}

\newcommand{\Poisson}[2]{\left\{#1 \,,\, #2\right\}}

\newcommand{\sign}{\text{sign}}

\newcommand{\fidmetric}[2]{\mathring{q}_{#1#2}}

\newcommand{\braopket}[3]{\left\langle \vphantom {#1 #2 #3} #1 \hphantom{|} \right| #2 \left| \hphantom{|} \vphantom {#1 #2 #3} #3 \right\rangle}

\begin{document}

\title{The Physical Relevance of the Fiducial Cell in Loop Quantum Cosmology}

\author{Fabio M. Mele}
\email{fabio.mele@oist.jp}
\affiliation{Okinawa Institute of Science and Technology Graduate University, 1919-1 Tancha, Onna-son, Okinawa 904-0495, Japan}

\author{Johannes M\"unch}
\email{johannes.muench@cpt.univ-mrs.fr}
\affiliation{Aix-Marseille Universit\'e, Universit\'e de Toulon, CNRS, CPT, 13288 Marseille, France}

%\begin{sffamily}

%\vspace{-0.25cm}

\begin{abstract}
%\noindent
A common way to avoid divergent integrals in homogeneous spatially non-compact gravitational systems, as e.g. cosmology, is to introduce a \textit{fiducial cell} by cutting-off the spatial slice at a finite region $V_o$.~This is usually considered as an auxiliary integral regulator to be removed after performing computations by sending $V_o \rightarrow \infty$.~In this paper, we analyse the dependence of the classical and quantum theory of homogeneous, isotropic and spatially flat cosmology on this fiducial cell.~We show that each fixed $V_o$ regularisation leads to a different canonically independent theory.~At the classical level, the dynamics of observables is not affected by the regularisation choice on-shell.~For the quantum theory, however, this leads to a family of regulator dependent quantum representations labelled by $V_o$ and the limit $V_o\to\infty$ becomes then more subtle.~First, we construct a novel isomorphism between different $V_o$-regularisations, which allows us to identify states in the different $V_o$-labelled Hilbert spaces to ensure equivalent dynamics for any value of the regulator.~The $V_o \rightarrow \infty$ limit would then correspond to choosing a state for which the volume assigned to the fiducial cell becomes infinite as appropriate in the late-time regime.~As second main result of our analysis, quantum fluctuations of observables smeared over subregions $V\subset V_o$, unlike those smeared over the full $V_o$, explicitly depend on the size of the fiducial cell through the ratio $V/V_o$ interpreted as the (inverse) number of subcells $V$ homogeneously patched together into $V_o$.~Physically relevant fluctuations for a finite region, as e.g. in the early-time regime, which would be unreasonably suppressed in a na\"ive $V_o\to\infty$ limit, become appreciable at small volumes.~Our results suggest that the fiducial cell is not playing the role of a mere regularisation but is physically relevant at the quantum level and complement previous statements in the literature.
%\vspace{0.75cm}
\end{abstract}
%\end{sffamily}

\maketitle

\section{Introduction}

Symmetry reduced models as cosmological or black hole spacetimes play an important role in the context of (canonical) quantum gravity approaches as e.g. loop quantum gravity (LQG) \cite{KieferQuantumGravity,ThiemannModernCanonicalQuantum,RovelliQuantumGravity}.~On the one hand, these models have a large amount of symmetries and are thus simple enough to allow for explicit quantum gravity computations.~On the other hand, they describe physically interesting systems containing gravitational singularities, whose understanding and ultimate fate in a theory surpassing general relativity are among the key questions that any quantum gravity theory aims to address (see~e.g.~\cite{BojowaldSingularitiesAndQuantum, NatsuumeTheSingularityProblem} and references therein).~Last but not least, the study of symmetry-reduced systems and their quantum aspects may open the way to potentially test possible phenomenological predictions of quantum gravity models (see e.g. \cite{AshtekarLoopquantumcosmology:astatusreport,AgulloLoopQuantumCosmology,AgulloLqcnon-Gaussianity,AgulloThePreInflationary} for a small sample of quantum cosmological models in the framework of loop quantum cosmology (LQC)).

However, a main conceptual and technical challenge lies in the fact that homogeneous gravitational systems such as cosmology with non-compact spatial topology lead to diverging actions and Hamiltonians.~In the LQC-literature \cite{AshtekarQuantumNatureOf,AshtekarAnIntroductionTo,AshtekarLoopquantumcosmology:astatusreport,AshtekarMathematicalStructureof,BojowaldQuantumcosmologyareview,BojowaldLoopQuantumCosmology,BodendorferAnElementaryIntroduction,BojowaldCriticalEvaluationof}, it is thus common to introduce a \textit{fiducial cell} to restrict the considered volume to a finite patch, labelled by the coordinate volume $V_o$.~The physical interpretation of such a fiducial cell has however been debated since then.~One point of view \cite{AshtekarQuantumNatureOf,AshtekarRobustnessOfKey,AgulloLoopQuantumCosmology,AshtekarLoopquantumcosmology:astatusreport} is to consider the fiducial cell as an infrared regulator and of purely auxiliary nature.~Physical results should not depend on the choice of the fiducial cell and the regulator can be then removed by sending $V_o \rightarrow \infty$ after Poisson brackets are evaluated to get back a non-compact space.~Another viewpoint based on effective quantum field theory (effective QFT) \cite{BojowaldCanonicalderivationof,BojowaldMinisuperspacemodelsas,BojowaldTheBKLscenario,BojowaldEffectiveFieldTheory,BojowaldCriticalEvaluationof} consists instead in interpreting it as the physical scale of homogeneity, which should evolve along with the evolution of the universe, thus being subject to an infrared renormalisation.~In particular, close to the classical singularity, this scale is actually microscopic \cite{BojowaldTheBKLscenario,BojowaldCriticalEvaluationof} due to the Belinskii-Khalatnikov-Lifshitz (BKL) scenario \cite{BelinskiiAgeneralsolution}.~From this point of view, the limit $V_o \rightarrow \infty$, if possible, should be related to a renormalisation group flow and the size of the fiducial cell acquires physical relevance as a renormalisation scale.~The question about how to interpret this fiducial structure has in turn profound consequences on the interpretation and validity of the \textit{effective approximation}, where the quantum physics is approximated by a smooth spacetime obeying effective modified dynamics.~The common belief in the literature is that the effective dynamics is applicable and quantum fluctuations are negligible even close to the resolved classical singularity \cite{TaverasCorrectionstothe,RovelliWhyAreThe,DienerNumericalsimulationsof,CorichiOnTheSemiclassical,SinghSemi-classicalStates} (see however \cite{BojowaldBlack-holemodels,BojowaldSpace-timephysicsin} for a recent review of further issues originating in effective models due to general covariance).

In this paper, we contribute to this discussion by analysing the role of the fiducial cell in detail.~We first recall the classical framework of cosmology and the necessity to introduce a fiducial cell at the off-shell level (Sec.~\ref{Sec:ClassCosmology}).~Physically relevant observables and their dependence on the regularisation are then discussed.~The classical part closes by arguing that indeed the classical physics, which means the dynamics of observables, is not altered by the choice of fiducial cell and the regulator can be removed by sending it to infinity at the on-shell level.~However, as the canonical structure depends on $V_o$, each cosmological theory regularised with a different fiducial cell is actually a canonically inequivalent theory.

This becomes relevant when moving to the quantum level in the second part of the paper (Sec.~\ref{Sec:QuantumCosmology}), where there is not a single quantum theory but rather a family of LQC quantum representations each corresponding to a different region $V_o$.~The $V_o$-rescaling symmetry of classical dynamics thus might or might not be broken at the quantum level.~Our main focus then lies on investigating how the quantum representations of differently regularised classical theories can be related.~More specifically, we ask: 1) how do we phrase the question about changing $V_o$ in the quantum theory so that it comes to be as close as possible to the situation at the classical level?~In other words, is it possible to implement a change of $V_o$, hence of the quantum theory, while keeping quantum dynamics the same?~2) what would then be the consequences for other quantum features such as uncertainty relations and fluctuations?~In Sec.~\ref{Sec:LQCquantisation}, we answer 1) in the affirmative by constructing a mapping between states in the different $V_o$-regularised Hilbert spaces with the same dynamics.~In Sec.~\ref{Sec:fluctuations} we then study the consequences for expectation values, higher statistical moments, and quantum fluctuations of both non-smeared and smeared operators.~In particular, the mapping between the different $V_o$-labelled Hilbert spaces allows us to study how these quantities scale in the different quantum representations associated to different fiducial cell sizes.~We find that moments and fluctuations of non-smeared elementary operators depend on the value of $V_o$, while those of the volume operator smeared over the entire fiducial cell do not, compatibly with previous work.~However, we further show that quantum fluctuations and uncertainty relations of canonically conjugate observables smeared over a subregion $V\subset V_o$ of the fiducial cell \emph{do} depend explicitly on the size of the fiducial cell through the ratio $V/V_o$ interpreted as the (inverse) number of subcells $V$ homogeneously patched into $V_o$.~Given then a small region $V$ that one is interested in tracking within a homogeneous model, its size can be operationally thought of as e.g.~given by the scale resolved by a detector.~Quantum fluctuations over the probe region $V$ become relevant in the early-time universe where also the size of the region $V_o$ over which homogeneity is imposed becomes small.~In the late-time universe, where homogeneity can be imposed over large scales, quantum fluctuations are instead suppressed the more subcells are patched together into a large region $V_o$.

Some concluding remarks and future directions are reported in Sec.~\ref{Sec:conclusion}.~More details will be worked out in a companion paper \cite{MeleOntheRoleof}, where a systematic symmetry reduction at the classical level and a full theory interpretation of the fiducial cell as the scale of the homogeneous truncation in field modes is also provided. 

\section{Classical Cosmology and the Fiducial Cell}\label{Sec:ClassCosmology}

We restrict ourselves to the simplest cosmological setting, that is homogeneous, isotropic, and spatially flat cosmology.
The metric
\begin{align}
	\dd s^2 &=-N(t)^2 \dd t^2 + q_{ab}\dd x^a\dd x^b\nonumber\\
	&= -N(t)^2 \dd t^2 + a(t)^2 \left(\dd x^2 + \dd y^2 + \dd z^2 \right)\;,\label{eq:FLRWmetric}
\end{align}

\noindent
is thus used as starting ansatz (see e.g. \cite{BodendorferAnElementaryIntroduction}).
In a classical treatment, it is possible to insert this ansatz into the Einstein equations coupled to a matter source and solve the differential equations for the scale factor $a(t)$, while the lapse $N(t)$ is a pure gauge choice and determines the interpretation of the coordinate $t$.

\begin{figure*}[!t]
%\begin{minipage}{\textwidth}
\centering
\begin{subfigure}[b]{0.375\textwidth}
\includegraphics[width=\textwidth]{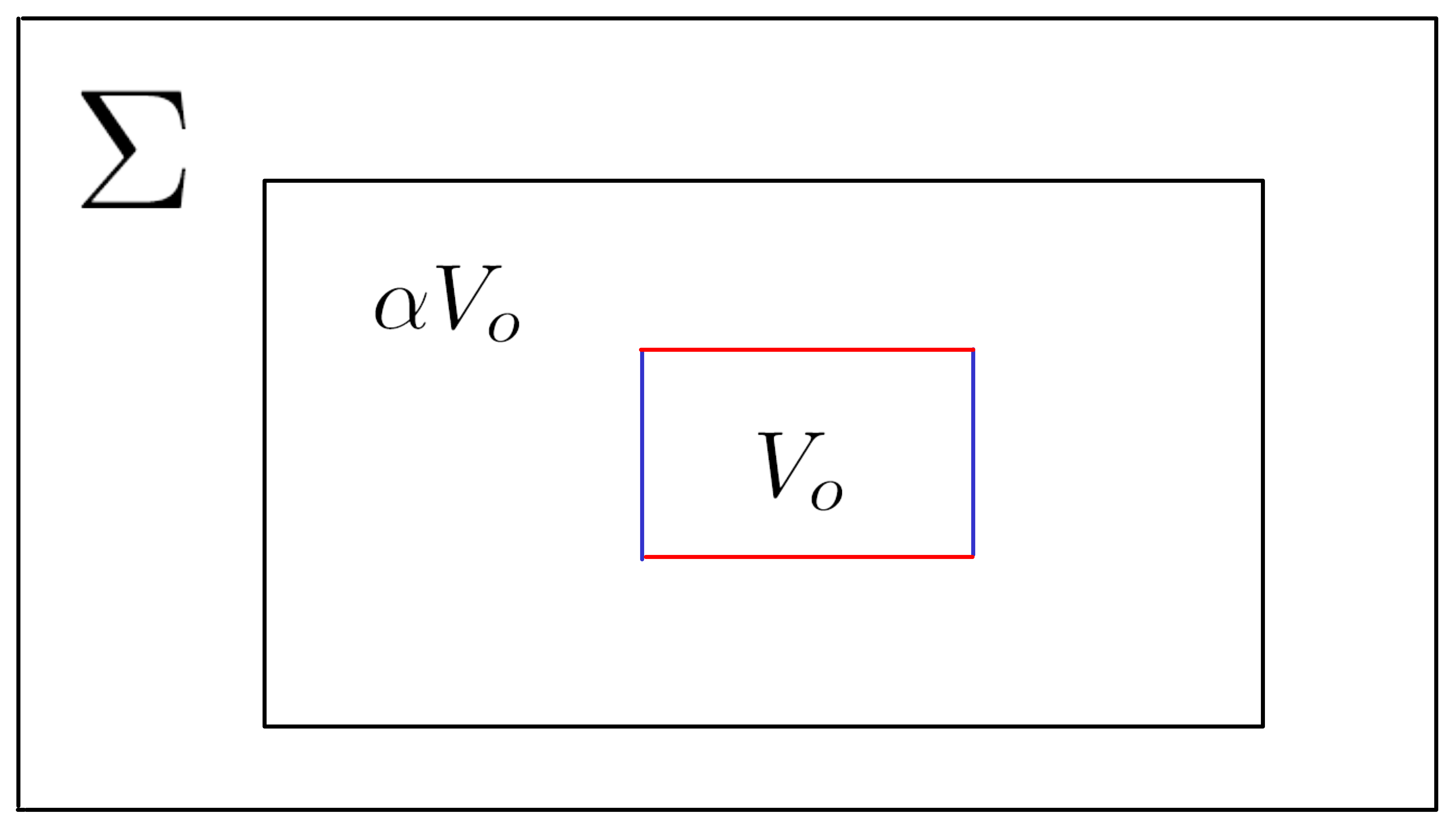}
\caption{}\label{Fig:subsystems1}
\end{subfigure}
\hspace{0.1\textwidth}
\begin{subfigure}[b]{0.375\textwidth}
\includegraphics[width=\textwidth]{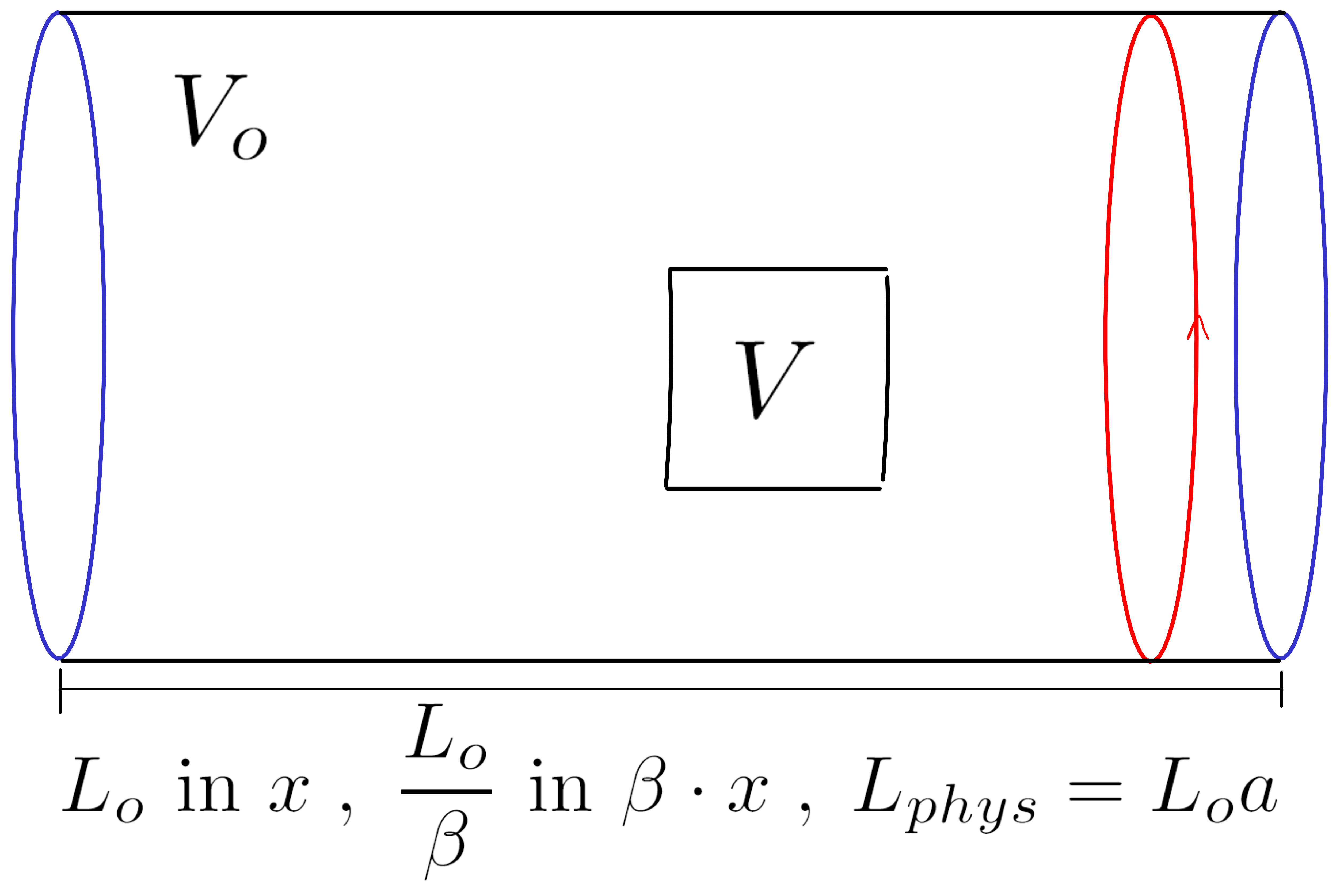}
\caption{}\label{Fig:subsystems2}
\end{subfigure}
\caption{(a) A compact fiducial cell $V_o\subset \Sigma$ of the non-compact spatial slice $\Sigma$ is chosen.
		It is equivalently possible to chose another, larger fiducial cell $V_o\subset\alpha V_o \subset \Sigma$, which is a coordinate independent choice of a larger subregion. 
		(b) To avoid boundary conditions the fiducial cell is equipped with a 3-torus topology, by identifying the blue lines and closing the red circle.
		The edge length of the fiducial cell $V_o$ and the circumference of the red circle are $L_o$ in the $x$-chart. This is coordinate dependent and it reads as $L_o/\beta$ in the $\beta x$-chart.
		The physical size is coordinate independent, but depends on initial conditions of $a$.
		On the spatial slice, we can consider finite volumes $V$, independent on the choice of $V_o$, as long as $V \subset V_o$..
}
\label{fig:sigma}
%\end{minipage}
\end{figure*}

In view of quantisation, it is however necessary to understand the off-shell structures such as action, Hamiltonian, and Poisson-structure first.
For this purpose, we fix the matter content of the universe to be a massless real scalar field $\phi$, which will play the role of physical clock later on.
Inserting the ansatz \eqref{eq:FLRWmetric} and $\phi(t,x) = \phi(t)$ into the Einstein-Hilbert action coupled to a massless scalar field, yields
\begin{align}
	\mathcal S &= \mathcal S_{EH} + \mathcal S_{M} = \int\dd t\,\mathcal L + \text{boundary terms} \;,\nonumber\\
	\mathcal L &= \int_{\Sigma} \dd^3 x\, \mathscr{L} = \int_{\Sigma}\dd^3x\, \left(-\frac{3}{\kappa} \frac{a\,\dot{a}^2}{N} + \frac{a^3 \dot{\phi}^2}{2N}\right)\label{action&lagrangian}
\end{align}

\noindent
where $\kappa = 8\pi G$ ($c =1$), dots refer to derivatives w.r.t. the $t$-coordinate, $\mathcal S_{EH} = \frac{1}{2\kappa} \int_\mathcal M\dd^4x \sqrt{g} R$ is the Einstein-Hilbert action and $\mathcal S_M = -\frac{1}{2}\int_\mathcal M \dd^4 x\sqrt{g} g^{\mu \nu}\partial_\mu \phi\partial_\nu \phi$ the matter action.
The four dimensional integral was split into a time-like integral and an integral over a spacial slice $\Sigma$. An immediate problem appears in the spatial integral, that is $\int_\Sigma \dd^3 x \rightarrow \infty$ as all fields are independent of the spatial point and thus do not satisfy proper fall off conditions.
The usual remedy provided in the literature (see e.g.  \cite{AshtekarQuantumNatureOf,AshtekarAnIntroductionTo,AshtekarLoopquantumcosmology:astatusreport,AshtekarMathematicalStructureof,BojowaldQuantumcosmologyareview,BojowaldLoopQuantumCosmology,BodendorferAnElementaryIntroduction,BojowaldCriticalEvaluationof}) is to restrict the spatial integral to a finite domain $V_o \subset \Sigma$ by introducing a so-called \textit{fiducial cell}, the latter being at this level a purely auxiliary topological construction to regularise the divergent integral.
The region $V_o$ is described by the coordinates $x,y,z \in \left[0,L_o\right]$ so that its coordinate volume\footnote{More precisely, the fiducial volume is computed w.r.t. a fiducial metric $\fidmetric{a}{b}$ whose coordinate axes are associated with the local triads along the edges of the cell, say $\fidmetric{a}{b}\dd x^a\dd x^b=\dd x^2_1+\dd x^2_2+\dd x^2_3$ (see e.g. \cite{AshtekarLoopquantumcosmology:astatusreport} for details). Moreover, to simplify the notation, we use the same notation $V_o$ for both the \textit{set} $V_o \subset \Sigma$ and its \textit{coordinate volume} $V_o = L_o^3$. Their distinction should be clear from the context.} is $\int_{V_o} \dd^3 x = L_o^3 =: V_o$.
An important question concerns now boundary conditions at $\partial V_o$, which arise when $\Sigma$ is assumed to be non-compact.
To avoid this problem, we topologically compactify $V_o$ by equipping it with a 3-torus topology $V_o \simeq_{top} \mathbb{T}^3$ so that $\partial V_o = \emptyset$ (cfr. Fig. \ref{fig:sigma}).

As this construction is purely auxiliary at first sight, it is usually expected to not enter any physical results in the end.
However, it should be kept in mind that already at this stage two main approximations have been done:
First, the boundaries are neglected here, i.e. we impose periodic boundary conditions, which amounts to ignore all the interactions of the region $V_o$ with the surrounding environment given by its complementary region in the spatial slice. 
Second, all other regions $\Sigma \setminus V_o$ are ignored and $V_o$ is simply extended by homogeneity. 
This excludes the possibility of changes in the spatial geometry from one fiducial cell to another, thus ignoring inhomogeneities with wavelengths larger than $L_o$. 
The latter can be safely neglected at large volumes, far from the bounce or classical singularity.
However, it remains to be examined how important these modes are to the total dynamics at small volumes.
Changing the size of $V_o$ makes in fact these approximations more or less restrictive.
In a follow up paper \cite{MeleOntheRoleof}, it will be presented how to do a systematic symmetry reduction at the Hamiltonian level by implementing spatial homogeneity via second-class constraints and using Dirac brackets.
In this framework, the fiducial cell can be interpreted as the scale on which the spacetime is actually homogeneous, which is thus a physical requirement.
Further, as we assume ${x,y,z} \in \left[0,L_o\right]^3$ and the torus topology here, this is the scale on which periodic boundary conditions are imposed.
Using the metric \eqref{eq:FLRWmetric}, this amounts to say that the spacetime is homogeneous and satisfies periodic boundary conditions on a length scale given by 
\begin{equation}\label{physicallength}
L_{phys} = \int_0^{L_o} \dd x \sqrt{g_{xx}} = L_o a(t) \;.
\end{equation}

\noindent
Its actual value thus depends on the solution and the initial conditions for $a(t)$.
The length $L_o$ then enters directly the physical quantity $L_{phys}$.
Let us emphasise here now an important difference between the coordinate size $L_o$ and the region $V_o\subset \Sigma$.
While a coordinate transformation $x \mapsto \beta x$ changes the value $L_o \mapsto L_o/\beta$, $L_{phys}$ is not changed as $a \mapsto \beta a$ \cite{AshtekarLoopquantumcosmology:astatusreport}.
The geometry of $V_o$ as described by a given solution $a(t)$ of the Einstein equations is not changed by this transformation, as this only amounts to describe it by means of different coordinates.
However, it is also possible to simply make the scale of homogeneity/periodicity larger by mapping $L_o \mapsto \alpha L_o$, which is not a coordinate transformation, but really physically rescales $V_o$ as a subset of $\Sigma$ (see Fig.~\ref{fig:sigma}), when referring to the same fixed solution\footnote{Note that this is a kinematical interpretation. Taking a closer look at the Einstein equations leads to the well-known fact that they are diffeomorphism invariant. Consequently, the metric with $a/\alpha$ has the same initial conditions for $L_{phys} = \alpha L_o a/\alpha$ and leads to equivalent physics.}.
In this case, we find $L_{phys} \mapsto \alpha L_{phys}$, which affects the physical length \eqref{physicallength}.
From now on, we keep the coordinates fixed and only consider transformations of the second kind, where the actual physical scale of periodicity, i.e. the physical volume of $V_o$, is changed.

In the following, we analyse the classical theory and its dependence on this fiducial cell and the topological compactification.
A Legendre transform of the Lagrangian \eqref{action&lagrangian}, restricted to $V_o$, leads to the Hamiltonian (see e.g. \cite{AshtekarLoopquantumcosmology:astatusreport,BodendorferAnElementaryIntroduction})
\begin{equation}
	H_T = V_o \cdot N \mathcal{H} \qquad , \qquad \mathcal H = -\frac{\kappa}{12}\frac{p_a^2}{a} + \frac{p_\phi^2}{2 a^3} \approx 0
	\end{equation}

\noindent
where the momenta are defined as $p_a = \partial \mathscr{L}/\partial \dot{a}$, $p_\phi = \partial \mathscr{L}/\partial \dot{\phi}$ with $\mathcal L = \int_{V_o} \dd^3x\,\mathscr{L} = V_o \mathscr{L}$, and satisfy the Poisson brackets
\begin{equation}
		\Poisson{a}{p_a} = \frac{1}{V_o} \qquad , \qquad \Poisson{\phi}{p_\phi} = \frac{1}{V_o} \;.
\end{equation}

\noindent
For the purposes of quantisation later, it is useful to perform the canonical transformation to the variables $v = a^3$ and $b =-p_a/(3 a^2)$, leading to
\begin{subequations}\label{eq:vb}
	\begin{equation}\label{eq:Hamiltonianvb}
		H_T = V_o\cdot N \mathcal{H} \qquad , \qquad \mathcal H = -\frac{3\kappa}{4} v b^2 + \frac{p_\phi^2}{2 v} \approx 0
	\end{equation}
	\begin{equation}\label{eq:poissonvb}
		\Poisson{b}{v} = \frac{1}{V_o} \qquad , \qquad \Poisson{\phi}{p_\phi} = \frac{1}{V_o} \;.
	\end{equation}
\end{subequations}

\noindent
The above $V_o$ factors can be understood from a field theory viewpoint where all the quantities involved in the above equations are given by the integrals over the spatial slice of the corresponding densities. More precisely, a systematic study of the reduction procedure \cite{MeleOntheRoleof} reveals that the canonical brackets \eqref{eq:poissonvb} can be actually understood as the Dirac brackets resulting from implementing homogeneity over the region $V_o$ via second-class constraints for the full theory variables. 

The interpretation of these variables becomes clear once smeared physical quantities are constructed.
In the full theory, the volume of a given region $V \subset V_o$ is given by
\begin{equation}\label{volumeofV}
	\text{vol}(V) = \int_V\dd^3 x \, \sqrt{q} = \int_V\dd^3 x \, v = V\cdot v\;.
\end{equation}

\noindent
Similarly, the variable $b$ can be interpreted as the rate at which (any) volume changes as\footnote{Comparing with \cite{BodendorferAnEmbeddingOf}, this variable has a full theory counterpart corresponding to $-\frac{1}{\text{vol}(V)} \int_{V} \dd^3 x q_{ab} P^{ab}$, with $P^{ab}$ the conjugate momentum to the spatial metric $q_{ab}$ which can be written in terms of the extrinsic curvature.}  
\begin{equation}
	\frac{1}{N \text{vol}(V)}\frac{\dd \text{vol}(V)}{\dd t} = \frac{1}{N \text{vol}(V)} \Poisson{\text{vol}(V)}{H_T} =\frac{3\kappa}{2} b \;,
\end{equation}

\noindent
which is independent of the particular volume $V$.
Similarly, the observable $p_\phi(V) = \int_V\dd^3 x p_{\phi} = Vp_\phi$ gives the total matter energy of the region $V \subset V_o$.
These are observables which have counterparts in the full theory and are independent of the choice of spatial coordinates. They reduce to the above simple expressions after imposing homogeneity and isotropy.
In the full theory they are independent for each possible choice of $V \subset V_o$, but after imposing homogeneity and isotropy, they are all related by a linear factor.
For instance, we have
\begin{equation}
	\text{vol}(V) = \frac{V}{V_o} \text{vol}(V_o)\;,
\end{equation}

\noindent
where $V/V_o$ is simply a topological ratio (i.e. independent of the metric) determining how many times $V$ fits into $V_o$.
The quantities $\text{vol}(V)$ and $p_\phi(V)$ are extensive as $\text{vol}(V_1) + \text{vol}(V_2) = \text{vol}(V_1 \cup V_2)$ for any disjoint regions $V_1$ and $V_2$ ($V_1 \cap V_2 = \emptyset$).
On the other hand, $b$ and $\phi$ are intensive and can be interpreted as averaged values of local quantities over the compact region $V_o$.
More details can be found in the follow up paper \cite{MeleOntheRoleof}.

Having this set up, we can study how the classical theory depends on the fiducial cell.
To this aim, let us consider a \textit{physical rescaling} $V_o \mapsto \alpha V_o$ (cfr. Fig.~\ref{fig:sigma}), i.e. $\text{vol}(V_o) = \int_{V_o}\dd^3 x \, v \mapsto \alpha \text{vol}(V_o)$.
Again, this is independent of coordinates and fixes the size of the fiducial torus $V_o$.
It is easy to see that the Hamiltonian \eqref{eq:Hamiltonianvb} is extensive and thus scales as
\begin{equation}\label{eq:Hscaling}
	H_T \longmapsto \alpha H_T \;.
\end{equation}

\noindent
Moreover, we have the following scaling behaviours
\begin{subequations}
	\begin{align}
		\text{vol}(V) \longmapsto \text{vol}(V) \quad &, \quad  \text{vol}(V_o) \longmapsto \alpha \text{vol}(V_o)\;,\label{eq:Vscaling}
		\\
		p_\phi(V) \longmapsto p_\phi(V) \quad &,\quad p_\phi(V_o) \longmapsto \alpha p_\phi(V_o)\;,
		\\
		b\longmapsto b \quad &,\quad \phi \longmapsto \phi \;,
	\end{align}
\end{subequations}

\noindent
from which we see that $\text{vol}(V)$, $p_\phi(V)$, $b$, and $\phi$ are independent of the size of the fiducial cell $V_o$.
Keeping this in mind, it becomes evident that the Poisson structure itself has to be dependent on $V_o$. Indeed, we have (cfr. Eqs. \eqref{eq:poissonvb} and \eqref{volumeofV}) 
\begin{align}
	\Poisson{b}{\text{vol}(V)} &= \frac{V}{V_o} \longmapsto \frac{1}{\alpha} \Poisson{b}{\text{vol}(V)}\;,\nonumber\\
	\Poisson{\phi}{p_\phi(V)} &= \frac{V}{V_o} \longmapsto \frac{1}{\alpha} \Poisson{\phi}{p_\phi(V)}\;,
\end{align}

\noindent
and, since the quantities entering the arguments of the Poisson bracket are independent of $V_o$, this scaling behaviour has to be a property of the bracket itself, namely
\be\label{eq:poissonscaling}
\Poisson{\;\cdot\;}{\;\cdot\;} \longmapsto \frac{1}{\alpha}\Poisson{\;\cdot\;}{\;\cdot\;}\;,
\ee

\noindent
where the arguments are also scaled. Summarising, changing the size of the scale of homogeneity (or in other words the periodic boundary conditions)  
\begin{itemize}
	\item is not a canonical transformation as the Poisson bracket and Hamiltonian are not preserved (the Hamiltonian rescales linearly, while the Poisson bracket rescales inversely linear);
	\item does not affect physical observables, that do not explicitly refer to the fiducial cell.
\end{itemize}

\noindent
Therefore, the choice of $V_o$ and its compactification affects the canonical theory.~Specifically, there are infinitely many canonically inequivalent cosmological theories all labelled by the value of $V_o = \int_{V_o} \dd^3 x$ and the choice of the fiducial cell $V_o \subset \Sigma$.~Still, the theories labelled by different values of the fiducial cell, say $V_o^{(1)}$ and $V_o^{(2)}$, can be related to each other in a well-defined manner.~In particular, it is possible to show that the dynamics of classical observables is independent of the specific value of $V_o$.~Indeed, let $H_T^{(1)}$, $H_T^{(2)}$, and $\Poisson{\;\cdot\;}{\;\cdot\;}_{(1)}$, $\Poisson{\;\cdot\;}{\;\cdot\;}_{(2)}$ be the Hamiltonians and Poisson brackets of the two theories with fiducial cells $V_o^{(1)}$, $V_o^{(2)}$, respectively.~For any observable $\mathcal{O}$, the time evolution is determined by
\begin{align}
	\dot{\mathcal O} &= \Poisson{\mathcal O}{H_T^{(1)}}_{(1)}\nonumber\\
	&=\frac{V_o^{(2)}}{V_o^{(1)}}\Poisson{\mathcal O}{\frac{V_o^{(1)}}{V_o^{(2)}}\cdot H_T^{(2)}}_{(2)}\nonumber\\
	&=\Poisson{\mathcal O}{H_T^{(2)}}_{(2)} \;, 
\end{align}

\noindent
which does not depend on the chosen value of $V_o$.
Explicitly, this can be seen as the Poisson bracket depends on $1/V_o$, while the Hamiltonian is linear in $V_o$ (cfr. Eqs.~\eqref{eq:vb}).

This result is not surprising, neither new \cite{AshtekarLoopquantumcosmology:astatusreport,AshtekarMathematicalStructureof,AshtekarQuantumNatureOf}, as we could have worked with the Einstein equations from the very beginning, which does not refer to $V_o$-dependent quantities such as an action, Hamiltonian or Poisson brackets.
The Einstein equations are local and as such do not depend on the global topology\footnote{Indirectly they do for a field theory due to boundary conditions, but these are already imposed by choosing a homogeneous metric ansatz.} so that no reference to the fiducial cell $V_o$ is needed.
In particular, it is possible to track the evolution of a finite volume $\text{vol}(V)$ and perform the limit $V_o \rightarrow \infty$ after evaluating the Poisson brackets.
As a classical theory only predicts dynamics, which is local, $V_o$ is really fiducial at the classical level and any reference to it can be removed afterwards by performing this limit (as required in e.g. \cite{AshtekarLoopquantumcosmology:astatusreport}), at least in principle, as long as the homogeneous approximations remains trustable\footnote{Some care is in fact still needed even at the classical level when one takes into account the fact that, after all, the homogeneous mini-superspace theory results from a mode truncation of dynamical fields in the full theory. From this perspective, the homogeneous description can be trusted as long as such a truncation provides us with a good approximation. In particular, as discussed in the follow up paper \cite{MeleOntheRoleof}, the fiducial cell sets the scale of a wavelength cut-off for the field modes and the error committed in the truncation of the inhomogeneous modes does depend on (inverse powers of) $V_o$.}.
However, in the next section, we will see how this situation changes in the case of a quantum theory.

\section{Implications for Quantum Cosmology}\label{Sec:QuantumCosmology}

\subsection{Quantisation of the Symmetry-Reduced Theory}\label{Sec:LQCquantisation}

Let us proceed now to quantise the family of classical theories labelled by $V_o$ using LQG techniques.
Note that the quantum theory can only be well defined for a finite $V_o$ as otherwise  the Poisson bracket, and thus the commutator, would be trivial. Consequently, it is unavoidable to construct a theory at finite $V_o$ and then examine the scaling behaviour of the relevant quantities.
As the classical Poisson bracket depends on the fiducial cell volume $V_o$ (see Eqs.~\eqref{eq:poissonvb},\eqref{eq:poissonscaling}) but the commutator cannot have this scaling property (it is just the composition of linear operators), each value of $V_o$ requires a separate quantum representation \cite{AshtekarMathematicalStructureof,AshtekarLoopquantumcosmology:astatusreport}, i.e. different Hilbert spaces and operators.

As a first step, the system is de-parametrised with the scalar field used as clock for the quantum dynamics.
The corresponding true Hamiltonian generating the evolution in $\phi$-time is obtained by solving the Hamiltonian constraint Eq.~\eqref{eq:Hamiltonianvb} for the canonical momentum $p_\phi(V_o)$ conjugate to $\phi$.
This leads to
\begin{equation}
	p_\phi(V_o) = \sqrt{\frac{3 \kappa}{2}} V_o v b =: H_{true}\;,
\end{equation}

\noindent
where we have only chosen the positive sign of the square root which is equivalent to consider only positive frequency modes.
This is a suitable simplification as the aim here is only to study the dependence on $V_o$. 
The considerations in the following straightforwardly extend to the case in which both positive and negative frequency modes are considered.

The next step consists of representing the operators $\hat{H}_{true}$, $\hat{v}$ and a regularisation of $b$ on the LQC-Hilbert space $\mathscr{H}_{LQC}$ given by \cite{AshtekarLoopquantumcosmology:astatusreport,AshtekarQuantumNatureOf,AshtekarMathematicalStructureof,BodendorferAnElementaryIntroduction}
\begin{subequations}
	\begin{align}
	&\mathscr{H}_{LQC} = L^2\left(\mathbb{R}_{Bohr},\dd\mu_{\text{Bohr}}\right) \quad, \quad \ket{\psi} = \sum_{\nu \in \mathbb R} \psi(\nu) \ket{\nu}\label{eq:LQCrepr1}
	\\
	&\braket{\nu|\nu'} = \delta_{\nu, \nu'} \quad,\quad \braket{b|\nu} = e^{i \lambda b \nu}  \quad,\quad \psi(\nu) = \braket{\nu|\psi}\label{eq:LQCrepr2}
\end{align}
\end{subequations}

\noindent
where $\ket{\nu}$ and $\ket{b}$ are respectively eigenstates of $\hat{v}$ and of $\widehat{e^{-i\lambda b}}$, see below.~Here, $\mathbb{R}_{Bohr}$ denotes the Bohr compactification of the real line \cite{AshtekarLoopquantumcosmology:astatusreport,AshtekarMathematicalStructureof,SubinDifferentialandPseudodiff}, a compact Abelian group which roughly speaking corresponds to the real line equipped with discrete topology.~Note the discrete sum in \eqref{eq:LQCrepr1} and the Kronecker-delta rather than a Dirac-delta normalisation in \eqref{eq:LQCrepr2}.~This is where the main difference between the weakly-discontinuous LQC polymer representation, motivated by the discreteness of spatial geometry at a fundamental scale, and the \emph{inequivalent} Schr\"odinger representation can be traced back \cite{AshtekarLoopquantumcosmology:astatusreport,AshtekarMathematicalStructureof,Ashtekar:shadowstates}.~As a consequence, only $\widehat{e^{-i\lambda b}}$ and not $\hat b$ comes to be a well-defined operator on the LQC Hilbert space in analogy with the use of holonomies in LQG\footnote{The classical variable $b$ is in fact directly related to the symmetry-reduced homogeneous and isotropic Ashtekar connection \cite{AshtekarLoopquantumcosmology:astatusreport,BojowaldLoopQuantumCosmology}.}.

The core issue is now the dependence of the quantum theory on $V_o$.~Different choices of $V_o$ correspond to different Hilbert spaces, although they are isomorphic and the difference is not explicitly visible.~It becomes visible in the operator representations. In fact, the $V_o$-dependence of the Poisson bracket between $v$ and $e^{-i\mu b}$ necessarily implies that the representations of the corresponding quantum operators will depend on it as well.~Specifically, the action of the elementary operators can be generically written as
\begin{equation}\label{reprelementaryoperators}
	\hat{v} \ket{\nu} = \frac{\eta^\gamma}{V_o^\gamma} \nu \ket{\nu} \quad , \quad \widehat{e^{-i\lambda \mu b}} \ket{\nu} = \ket{\nu -\frac{\lambda \mu}{\eta^\gamma V_o^\delta}} \;,
\end{equation}

\noindent
with arbitrary powers $\gamma,\delta$ such that $\gamma + \delta = 1$, $\eta  = \kappa^{3/2}$, and units $\hbar = 1$, $[\kappa]  = length^2$, $[\nu] = [\hat{v}] = [\mu] = 1$, $[b] = [\lambda^{-1}] = length^{-3}$. $\mu \in \mathbb{R}$ is an arbitrary dimensionless number and $\lambda$ is the so-called polymerisation scale. The powers of $\gamma$ and $\delta$ are just a notation we introduced to reflect the freedom in incorporating the $V_o$ factors in the representation of the quantum operators to ensure the Poisson brackets to be correctly represented as commutation relations.~The $V_o$-factors in \eqref{reprelementaryoperators} are in fact unavoidable for the commutator to match the correspondence principle with the classical Poisson bracket, i.e.
\be
\left[\widehat{e^{-i\lambda \mu b}}, \hat{v}\right]\ket{\nu}=\frac{\lambda\mu}{V_o}\widehat{e^{-i\lambda \mu b}}\ket{\nu}\quad,\quad\forall\,\ket{\nu}
\ee
so that (cfr.~\eqref{eq:poissonvb})
\be\label{correspondenceprinciple}
\left[\widehat{e^{-i\lambda \mu b}}, \hat{v}\right] = i\, \reallywidehat{\Poisson{e^{-i\lambda \mu b}}{v}} \;.
\ee

\noindent
At the quantum level, the scaling property of the Poisson bracket (cfr. Eq.~\eqref{eq:poissonscaling}) is thus shifted into the quantisation map for the representations of the operators associated to phase space quantities.
Consequently, although the classical phase space function $v$ is independent of $V_o$, its quantum representation does depend on it and, for any two distinct values $V_o^{(1)}$ and $V_o^{(2)}$, we have
\be\label{eq:Palphaopscaling}
\hat{v}\Bigl|_{V_o^{(1)}} = \left(\frac{V_o^{(2)}}{V_o^{(1)}}\right)^{\gamma}\hat{v}\Bigr|_{V_o^{(2)}}\;,
\ee

\noindent
and
$$
\widehat{e^{-i\lambda\mu^{(1)} b}}\Bigl|_{V_o^{(1)}} = \widehat{e^{-i \lambda \mu^{(2)}b}}\Bigl|_{V_o^{(2)}} \quad,\quad \mu^{(1)} = \left(\frac{V_o^{(2)}}{V_o^{(1)}}\right)^\delta \mu^{(2)}.
$$

\noindent
Finally, standard LQC ordering and regularisation prescriptions for the Hamiltonian \cite{Martin-BenitoFurtherImprovementsIn} lead to
\begin{align}
	\hat{H}_{true} &=\sqrt{\frac{3 \kappa}{2}} V_o \sqrt{|\hat{v}|} \Biggl(\frac{\widehat{\sin\left(\lambda b\right)}}{2\lambda} \sign\left(\hat{v}\right)\nonumber\\
 &\qquad+\sign\left(\hat{v}\right) \frac{\widehat{\sin\left(\lambda b\right)}}{2\lambda}\Biggr) \sqrt{|\hat{v}|} \;,\label{eq:MMOHtrue}
\end{align}

\noindent
whose action, after defining 
\begin{equation}\label{eq:defalphatheta}
	\nu = \frac{\lambda}{\eta^\gamma V_o^\delta} n \quad , \quad n \in \mathbb{R} \quad , \quad \theta = \frac{\lambda}{\eta^\gamma V_o^\delta} \;,
\end{equation}

\noindent
can be recast into the following finite difference equation
\begin{align}
	\hat{H}_{true} \psi(\nu) &=\braopket{\nu}{\hat{H}_{true}}{\psi}
	\notag
	\\
	&= \frac{i}{4} \sqrt{\frac{3 \kappa}{2}}\Bigl(s_+(n)\,\sqrt{|n||n+1|} \psi\left(\theta\cdot(n+1)\right)\nonumber\\
	&\quad- s_-(n)\,\sqrt{|n||n-1|} \psi\left(\theta\cdot(n-1)\right)\Bigr)\;.\label{Htrueaction}
\end{align} 

\noindent
with $s_\pm(n)= \sign(n\pm1) + \sign(n)$.~As usual the zero volume state $\nu = 0$ is annihilated and positive and negative branches are preserved.~The quantity $\theta$ entering Eq.~\eqref{Htrueaction} can be thought of as an ``effective'' polymerisation scale, which is explicitly $V_o$-dependent (cfr.~Eq.~\eqref{eq:defalphatheta}).~Moreover, we note that the linear scaling behaviour of the classical Hamiltonian (cfr.~Eq.~\eqref{eq:Hscaling}) is again changed by the $V_o$-quantum representation and is rather undetermined.~However, to have a well-defined dynamics given by the Sch\"odinger equation generated by $\hat{H}_{true}$
\begin{equation}
	i\partdif{}{\phi} \ket{\psi} = \hat{H}_{true} \ket{\psi} \quad , \quad \ket{\psi ; \phi} = e^{i\phi \hat{H}_{true}} \ket{\psi} \;, 
\end{equation}

\noindent
such a scaling behaviour has to be determined.
In this respect, it is possible to identify states in the different $V_o$-quantum theories which have the same evolution behaviour.
Let us emphasise that, in principle, it might be possible to allow $V_o$ dependent dynamics and forget about the following identification.
However, as the classical dynamics is independent of $V_o$, it seems reasonable to demand the same for quantum dynamics.

For this construction, we make use of the fact that there exists \cite{Martin-BenitoFurtherImprovementsIn,AshtekarCastinglqcinthe} an analytic expression for a function $\Psi_E: \mathbb{R} \rightarrow \mathbb{C}$ satisfying
\begin{widetext}
\begin{equation}\label{eigenstatesHtrue}
	-\frac{i}{2} \sqrt{\frac{3 \kappa}{2}} \cdot \left(s_+(n)\,\sqrt{|n||n+1|} \Psi_E\left(n+1\right) - s_-(n)\,\sqrt{|n||n-1|} \Psi_E\left(n-1\right)\right) = E \Psi_E(n) \;.
\end{equation}
\end{widetext}

\noindent
Consequently, the states $\psi_E(\nu) = \psi_E(\theta\cdot n) = \Psi_E(n)$ are eigenstates of the Hamiltonian \eqref{Htrueaction} with eigenvalue $E$.~As shown in \cite{Martin-BenitoFurtherImprovementsIn}, the operator \eqref{eq:MMOHtrue} is essentially self-adjoint with absolutely continuous, non-degenerate spectrum equal to the real line.~The explicit expression of the corresponding eigenfunctions can be found in \cite{Martin-BenitoFurtherImprovementsIn,AshtekarCastinglqcinthe} but is not relevant for the considerations that follow.~Consider now two quantum representations resulting from quantising the classical theories associated with two distinct values of $V_o$, say $V_o^{(1)}$ and $V_o^{(2)}$, respectively.
In both representations one could find the eigenstates of the corresponding Hamiltonian according to the same identification \eqref{eigenstatesHtrue} ($V_o$ and thus $\theta$ are different).
It follows that in both quantum representations one finds the same spectrum for $\hat{H}_{true}$ and, recalling Eq. \eqref{eq:defalphatheta}, the eigenstates can be related by
\begin{equation}\label{eq:eigenstatesmapping}
	\psi_E^{(1)}\left(\nu\right) = \psi_E^{(2)}\left(\left(\frac{V_o^{(1)}}{V_o^{(2)}}\right)^{\delta}\nu\right) \;,
\end{equation}

\noindent
where the super-scripts ${(1)}$ and $(2)$ denote the eigenstates in the corresponding $V_o$-quantum theory.
Let us stress that a priori the carrier Hilbert spaces for the different quantum representations are \textit{different} and it is only due to the simplicity of the model under consideration that they can be easily identified as above.~As anticipated above, it is a priori not necessary to identify the states in the different quantum theories as in \eqref{eq:eigenstatesmapping} and, depending on the aims and/or the physical situation to be described, one could in principle find different arguments leading to a different identification.~Here, the relation \eqref{eq:eigenstatesmapping} is to ensure the dynamics to be the same in the two quantum theories, similar to the classical case. To see this, we note that the above identification of eigenstates allows to define a well-defined transformation behaviour of the Hamiltonian operator according to
\begin{widetext}
\begin{equation}
	\left.\hat{H}_{true}\right|_{V_o^{(1)}} = \int \dd E\, E \braket{\psi_E^{(1)}|\;\cdot\;} \psi_E^{(1)} \quad\longmapsto\quad \left.\hat{H}_{true}\right|_{V_o^{(2)}} = \int \dd E\, E \braket{\psi_E^{(2)}|\;\cdot\;} \psi_E^{(2)}\;,
\end{equation}
\end{widetext}

\noindent
from which it is evident that two states $\psi^{(1)} \in \mathscr H_{LQC}^{(1)}$ and $\psi^{(2)} \in \mathscr H_{LQC}^{(2)}$ have the exact same $\phi$-time evolution as long as $\braket{\psi_E^{(1)}|\psi^{(1)}}_{(1)} = \braket{\psi_E^{(2)}|\psi^{(2)}}_{(2)}$ for all $E$.
This in turn induces, under the demanding of dynamics to remain $V_o$-independent, the following dynamics-preserving isomorphism between the two different quantum theories
\begin{equation*}
	\mathscr{I}:\; \mathscr{H}_{LQC}^{(1)} \longrightarrow \mathscr{H}_{LQC}^{(2)}
\end{equation*}
by
\begin{align}
\psi^{(1)} \longmapsto \,&\psi^{(2)}= \mathscr{I}\left(\psi^{(1)}\right)
	\notag
	\\
	&\psi^{(2)}(\nu) = \psi^{(1)}\left(\left(\frac{V_o^{(2)}}{V_o^{(1)}}\right)^\delta \nu\right) \;.\label{eq:isomorphygrav}
\end{align}

\noindent
Therefore, it is possible to make the whole families of quantum theories labelled by the values of $V_o$ dynamically equivalent, as it was the case for the classical theory.
However, a quantum theory is more than just dynamics, but also includes quantum uncertainty relations and fluctuations.
Their dependence on $V_o$ is discussed in the next subsection.~In this respect, let us notice that the above isomorphism implementing a notion of fiducial cell rescaling at the quantum level is in fact the sought mapping e.g.~in \cite{CorichiOnTheSemiclassical} and, as we shall see in the next subsection, can be used for a detailed analysis of expectation values and higher moments (see Eq.~\eqref{eq:scalingmomentsGR} below). The present results can in particular be used to study the scaling properties of semiclassical states as e.g.~those considered in \cite{CorichiOnTheSemiclassical,CorichiCoherentSemiclassicalStates,FraserTamingFluctuationsfor}. We refer the interested reader to the companion paper \cite{MeleOntheRoleof} for details.

\subsection{Uncertainty Relations and Quantum Fluctuations}\label{Sec:fluctuations}

Having determined the mapping \eqref{eq:isomorphygrav} between the Hilbert spaces associated with different $V_o$-valued quantum representations, we can now study whether and how expectation values and higher statistical moments change from one $V_o$ value to the other.~Note that all the expectation values in what follows are $\phi$-time dependent and, since the isomorphism preserves time evolution, the following statements hold in a fully dynamical sense.~However, to ease the notation, we shall suppress the explicit $\phi$-time dependence and write $\psi(\phi) = \psi$, $\ket{\psi; \phi} = \ket{\psi}$.

As for the operator $\hat{v}$, recalling the action Eq.~\eqref{reprelementaryoperators} together with the definitions Eq.~\eqref{eq:defalphatheta}, we find
\begin{widetext}
\begin{align}
	\left<\left.\hat{v}\right|_{V_o^{(1)}}\right>_{\psi^{(1)}}&:=\braopket{\psi^{(1)}}{\left.\hat{v}\right|_{V_o^{(1)}}}{\psi^{(1)}} = \sum_{\nu \in \mathbb R} \psi^{(1)*}(\nu) \frac{\eta^\gamma}{\left(V_o^{(1)}\right)^\gamma} \nu \psi^{(1)}(\nu)
	\notag
	\\
	&=\sum_{\nu \in \mathbb R} \psi^{(2)*}\left(\left(\frac{V_o^{(1)}}{V_o^{(2)}}\right)^\delta\nu\right) \frac{\eta^\gamma}{\left(V_o^{(1)}\right)^\gamma} \nu \psi^{(2)}\left(\left(\frac{V_o^{(1)}}{V_o^{(2)}}\right)^\delta\nu\right)
	\notag
	\\
	&= \frac{V_o^{(2)}}{V_o^{(1)}}\sum_{\nu' \in \left(\frac{V_o^{(1)}}{V_o^{(2)}}\right)^\delta \mathbb R} \psi^{(2)*}\left(\nu'\right) \frac{\eta^\gamma}{\left(V_o^{(2)}\right)^\gamma} \nu' \psi^{(2)}\left(\nu'\right)
	\notag
	\\
	&= \frac{V_o^{(2)}}{V_o^{(1)}} \left<\left.\hat{v}\right|_{V_o^{(2)}}\right>_{\psi^{(2)}} \;,
\end{align}
\end{widetext}

\noindent
where $\nu' = \left(V_o^{(1)}/V_o^{(2)}\right)^\delta\nu$ in the second to last line.
By construction of \eqref{eq:isomorphygrav}, we further have that
\begin{equation}\label{eq:qHscaling}
	\left<\left.\hat{H}_{true}\right|_{V_o^{(1)}}\right>_{\psi^{(1)}} = 
	\left<\left.\hat{H}_{true}\right|_{V_o^{(2)}}\right>_{\psi^{(2)}} \;, 
\end{equation}

\noindent
and it is straightforward to verify the following relations
\begin{subequations}\label{eq:scalingmomentsGR}
	\begin{align}
		\left<\left.\hat{v}\right|_{V_o^{(1)}}^n\right>_{\psi^{(1)}}&=\; \left(\frac{V_o^{(2)}}{V_o^{(1)}}\right)^n \left<\left.\hat{v}\right|_{V_o^{(2)}}^n\right>_{\psi^{(2)}}\;,\\
		\left<\left.\widehat{e^{-i\lambda \mu b}}\right|_{V_o^{(1)}}^n\right>_{\psi^{(1)}} &=\;  \left<\left.\widehat{e^{-i\lambda \mu b}}\right|_{V_o^{(2)}}^n\right>_{\psi^{(2)}}\;,\\
		\Delta_{\psi^{(1)}} \left.\hat{v}\right|_{V_o^{(1)}} &=\; \frac{V_o^{(2)}}{V_o^{(1)}} \Delta_{\psi^{(2)}} \left.\hat{v}\right|_{V_o^{(2)}}\;,\\
		\Delta_{\psi^{(1)}} \left.\widehat{e^{-i\lambda \mu b}}\right|_{V_o^{(1)}} &=\; \Delta_{\psi^{(2)}} \left.\widehat{e^{-i\lambda \mu b}}\right|_{V_o^{(2)}}\;,
	\end{align}
\end{subequations}

\noindent
for the higher moments and variances of the operators in Eq.~\eqref{reprelementaryoperators}.
With this transformation behaviour at our disposal, we can then investigate possible physical effects on the observables
\begin{equation}\label{intvolumeoperators}
	\widehat{\text{vol}\left(V_o\right)} = V_o \hat{v}\quad ,\quad \widehat{\text{vol}\left(V\right)} = V \hat{v} = \frac{V}{V_o} \widehat{\text{vol}\left(V_o\right)} \;,
\end{equation}

\noindent
corresponding to the integrated volume of the fiducial cell $V_o \subset \Sigma$ or any region $V\subseteq V_o$ (see Fig.~\ref{fig:sigma}).
According to the results in Eqs.~\eqref{eq:scalingmomentsGR}, we have
\be\label{expvalueintvolume}
	\left<\left.\widehat{\text{vol}\left(V_o^{(1)}\right)}\right|_{V_o^{(1)}}\right>_{\psi^{(1)}} = \left<\left.\widehat{\text{vol}\left(V_o^{(2)}\right)}\right|_{V_o^{(2)}}\right>_{\psi^{(2)}}
\ee

\noindent
i.e., in contrast to the classical theory (cfr. Eq.~\eqref{eq:Vscaling}), the expectation value of the operator for the integrated volume of the cell is independent of the value of its coordinate volume.
Note that changing $V_o$ actively changes the subset in the spatial slice $\Sigma$ and the region on which periodic boundary conditions are imposed.

Similarly, using the relations \eqref{correspondenceprinciple}, \eqref{eq:scalingmomentsGR}, and \eqref{expvalueintvolume}, the uncertainty relations read as\footnote{To ease the comparison with existing literature as e.g. \cite{TaverasCorrectionstothe,RovelliWhyAreThe}, we focus here on the operator corresponding to the simplest regularisation for the conjugate momentum $b$ by combination of point holonomies $e^{\pm i\lambda b}$ yielding the sin function.}
\begin{widetext}
\begin{align}
		\frac{1}{2} \left|\left<\left.\widehat{\cos\left(\lambda b\right)}\right|_{V_o^{(1)}}\right>_{\psi^{(1)}}\right| &\le \Delta_{\psi^{(1)}} \left.\widehat{\text{vol} \left(V_o^{(1)}\right)}\right|_{V_o^{(1)}} \Delta_{\psi^{(1)}} \left.\widehat{\frac{\sin\left(\lambda b\right)}{\lambda}} \right|_{V_o^{(1)}}
		\notag
		\\
		&= \Delta_{\psi^{(2)}} \left.\widehat{\text{vol}\left(V_o^{(2)}\right)}\right|_{V_o^{(2)}} \Delta_{\psi^{(2)}} \left.\widehat{\frac{\sin\left(\lambda b\right)}{\lambda}} \right|_{V_o^{(2)}} \ge \frac{1}{2} \left|\left<\left.\widehat{\cos\left(\lambda b\right)}\right|_{V_o^{(2)}}\right>_{\psi^{(2)}}\right| \;.\label{volVouncertainty}
\end{align}
\end{widetext}

\noindent
Therefore, the expectation values and fluctuations of the elementary operators depend on the value of $V_o$ as per Eqs.~\eqref{eq:scalingmomentsGR}, while those for the observable measuring the size of the full fiducial cell are independent of the specific value of $V_o$.~Independently of which subset $V_o \subset \Sigma$ is chosen, the expectation values and uncertainty relations involving the volume operator smeared over the entire $V_o$ are only state dependent.~Moreover, if the state $\psi^{(1)}$ saturates the uncertainty relation, then $\psi^{(2)}$ does too.~Similar is true for the energy density
\be\label{eq:energydensity}
\rho_\psi(\phi):= \frac{\left<\widehat{p_\phi(V_o)}\right>_\psi^2}{\left<\widehat{\text{vol}(V_o)}\right>_\psi^2} = \frac{\left<\hat{H}_{true}\right>_\psi^2}{\left<\widehat{\text{vol}(V_o)}\right>_\psi^2} \;,
\ee
\noindent
which due to Eqs.~\eqref{eq:Hscaling}, \eqref{expvalueintvolume} is independent of the $V_o$-representation.
Further, as the isomorphism Eq.~\eqref{eq:isomorphygrav} preserves dynamics, the energy $1/2\lambda^2$ at the bounce is independent, too. This works also for a subregion $V\subset V_o$ as both quantities at the numerator and denominator of \eqref{eq:energydensity} are extensive so that the multiplicative factors would cancel yielding an intensive ratio.

It is then insightful to evaluate the expectation value of the total volume for an eigenstate of $\hat{v}$, i.e. $\psi = \delta_{\nu,\nu_o}$, which reads as

\be
\left<\widehat{\text{vol}\left(V_o\right)} \right>_{\delta_{\nu,\nu_o}} = V_o \frac{\eta^\gamma}{V_o^\gamma} \nu_o \overset{\gamma+\delta=1}{=} V_o^\delta\eta^\gamma\nu_o\overset{\eqref{eq:defalphatheta}}{=} \lambda n_o \;.
\ee

\noindent
Consequently, for any choice of $V_o$, it is only possible to assign to this region the volume $\lambda n_o$, where $n_o \in \mathbb{N}$ after imposing the Hamiltonian constraint (and restricting to positive volumes).~Compatibly with \eqref{expvalueintvolume}, this is obviously independent of $V_o$, i.e. the (topological) scale on which homogeneity and periodicity are imposed.~This is plausible as $V_o^{(1)}$ or $V_o^{(2)}$ are purely topological constructions and there is no reference to any geometry.~The geometry in turn is then specified by the choice of state $\psi$, which is in a sense the quantum equivalent of specifying initial conditions for the classical (coordinate-independent) observable $\text{vol}(V_o)$.~As there is no additional reference field in the system, there is no way to distinguish $V_o^{(1)}$ from $V_o^{(2)}$.

The situation is however different for a finite region $V \subset V_o^{(1)}, V_o^{(2)} \subset \Sigma$ as can be seen by studying the change of physical volume assigned to it and relating it to the total fiducial cell.~As discussed in Sec.~\ref{Sec:ClassCosmology}, changing $V_o$ in the classical theory is a symmetry of the dynamics and the limit $V_o \rightarrow \infty$, $\text{vol}(V_o) \rightarrow \infty$ can be taken for any finite initial conditions on $v$\footnote{Let us remind that, even though the dynamics of classical observables is not affected in the $V_o\to\infty$ limit, such a limit would spoil the canonical structure of the classical theory (cfr. Eq. \eqref{eq:poissonvb}). A na\"ive limit which ignores the fact that different quantum representations and Hilbert spaces are identified by the different values of $V_o$, would in turn spoil the (off-shell) starting point for the canonical commutation relations.}.~The quantum case requires more care.~To this aim, let us study the scaling behaviour of the expectation values of the operator $\widehat{\text{vol}(V)}$ and the corresponding uncertainty relations.~These can be readily computed to be (for a given value of $V_o$)
\begin{align}
    &\left<\widehat{\text{vol}(V)}\right>_\psi =  \frac{V}{V_o}\left<\widehat{\text{vol}(V_o)}\right>_\psi\;,\\
    &\Delta_\psi \widehat{\text{vol}(V)} \Delta_\psi \frac{\widehat{\sin\left(\lambda b\right)}}{\lambda} \ge \frac{V}{2\,V_o} \left|\left<\widehat{\cos\left(\lambda b\right)}\right>_\psi\right|\;,\label{eq:DeltavolV}
\end{align}
 from which we have the following relations between the quantum theories corresponding to two different values $V_o^{(1)}$ and $V_o^{(2)}$ (cfr.~\eqref{intvolumeoperators},~\eqref{expvalueintvolume}, and \eqref{eq:scalingmomentsGR})
\begin{align}
 &\left<\widehat{\text{vol}(V)}\right>_{\psi^{(2)}} =  \frac{V_o^{(1)}}{V_o^{(2)}}\left<\widehat{\text{vol}(V)}\right>_{\psi^{(1)}}\;,\label{volVscalingexpvalue}\\
&\Delta_{\psi^{(2)}} \widehat{\text{vol} \left(V\right)} \Delta_{\psi^{(2)}} \widehat{\frac{\sin\left(\lambda b\right)}{\lambda}}= \frac{V_o^{(1)}}{V_o^{(2)}}\Delta_{\psi^{(1)}} \widehat{\text{vol}\left(V\right)} \Delta_{\psi^{(1)}} \widehat{\frac{\sin\left(\lambda b\right)}{\lambda}}.\label{volVscalinguncert}
\end{align}
Thus, not only the expectation value and fluctuations of the elementary operator $\hat v$ in \eqref{eq:scalingmomentsGR} but also those for the volume of $V$ depend on the value of $V_o$.~This is to be contrasted with Eqs.~\eqref{expvalueintvolume},~\eqref{volVouncertainty} for the entire fiducial cell which have been the focus of previous work.~It is then worth noticing that Eq.~\eqref{eq:DeltavolV} extends previous results in the literature to the case in which one considers a sub-region $V$ of the fiducial cell $V_o$ and, consistently, previous results are recovered for $V=V_o$ (cfr. Eqs.~(20) and (34) in \cite{RovelliWhyAreThe}).

Let us discuss the above $V_o$-dependence in more detail.~The quantum dynamics of the volume $\widehat{\text{vol}\left(V\right)}$ of the finite region $V$ is not affected either by the coordinate volume $V_o$ or the \textit{physical volume} assigned to the fiducial cell $V_o$ as long as 
\begin{equation}\label{volVdoublelimit}
	\left<\widehat{\text{vol}(V)}\right>_\psi =  \frac{V}{V_o}\left<\widehat{\text{vol}(V_o)}\right>_\psi \xrightarrow{\frac{V}{V_o} \rightarrow 0\,, \,\left<\widehat{\text{vol}(V_o)}\right>_\psi \rightarrow \infty} \;\text{finite}.
\end{equation}

\noindent
The quantity \eqref{volVdoublelimit} might be a proper cosmological observable as the double scaling limit might be performed such that $\left<\widehat{\text{vol}(V)}\right>_\psi \sim$ \textit{size of universe today}.~Consequently, the state $\psi$ and the observable $\widehat{\text{vol}(V)}$ provide a physically reasonable description of the volume of the universe.~Now, as the (expectation value of the) total volume of the region $V_o$ in a given state does not depend on the value of $V_o$ (cfr. Eq.~\eqref{expvalueintvolume}), the quantum counterpart of a putative classical $V_o\to\infty$ limit would be to chose a state for which $\left<\widehat{\text{vol}(V_o)}\right>_{\psi} \rightarrow \infty$, i.e. the region of homogeneity is enlarged geometrically.~This is the case in late time cosmology where the homogeneous approximation can be safely trusted on large scales and, for finite $V$ and sufficiently large $V_o$, quantum fluctuations \eqref{eq:DeltavolV} are suppressed by the ratio $V/V_o$ and remain so over different, yet large sizes of $V_o$ (cfr.~\eqref{volVscalinguncert}).

The situation is rather different at the small scales of the early-time universe where the volume of both $V$ and $V_o$ become sufficiently small so that quantum fluctuations cannot be neglected.~The importance of fluctuations for finite small cells has been already emphasised e.g.~in \cite{BojowaldTheBKLscenario,BojowaldCriticalEvaluationof}.~Evolving the system backward following the collapse of an initially large-scale homogeneous universe, structure forms and inhomogeneities build up within a co-moving volume of given coordinate size $V_o$.~For the collapse process to be still described using a homogeneous model, a smaller region over which homogeneity is imposed should be selected when inhomogeneities become appreciable so that more and more inhomogeneous modes with wavelength larger than the fiducial cell can be neglected within that region.~This would be in particular required to be eventually consistent with an asymptotic BKL scenario according to which spacetime dynamics is locally homogeneous when approaching the space-like singularity.~From the scaling property \eqref{volVscalinguncert} we thus see that, in the early-time regime where $V_o^{(1)}\sim V_o^{(2)}\sim V_o$ are comparable and small, not only the quantum fluctuations for the elementary operators \eqref{eq:scalingmomentsGR} but also those for the smeared volume of a small region $V\subset V_o$ are not suppressed by the $V/V_o$ factor in \eqref{eq:DeltavolV}.~The states yielding small physical volumes are thus very quantum and the quantum
description of $V$ at small scales cannot be made effectively classical to an arbitrary precision.

Given the region $V$ one is interested in describing within the homogeneous approximation as e.g.~a small elementary cell inside $V_o$, and thinking of the latter as been made of multiple identical elementary cells patched together homogeneously\footnote{As discussed in the companion paper \cite{MeleOntheRoleof}, this is precisely the picture one gets by explicitly carrying out the homogeneous reduction of the gravitational field modes where the spatial slice is partitioned into disjoint cells and homogeneity is imposed only on a finite number of them $V_o=\bigsqcup_n V_n$.}, quantum fluctuations of the individual cells are present at small scales and it is only when going from the early-time small scales to the late-time large scales that they are suppressed as the number $V_o/V$ of subcells fitting into a bigger and bigger $V_o$ grows\footnote{This is also in line with some of the discussions about coarse graining and renormalisation with coherent states in LQC \cite{BodendorferCoarseGrainingAs,BodendorferRenormalisationwithsu11} according to which, by patching together $N$ cells, the fluctuations in the large cell grow only as $\sqrt{N}$, so that the relative fluctuations are vanishing in the infinite cell limit.}.~As emphasised in Sec.~\ref{Sec:LQCquantisation}, the mapping between the different $V_o$-labelled quantum theories is a priori not unique.~Therefore, one could in principle seek for a different mapping for which the resulting scaling behaviours are such that quantum fluctuations remain small when shrinking $V_o$.~However, if such a mapping exists, the dynamics of the states will be modified under a change of the region $V_o$.~In either case, it is clear that the size of the fiducial cell has a physical significance and is not just a regulator that can be removed at the end of the day with no effects at the quantum level.

\section{Conclusions}\label{Sec:conclusion}

In this paper, we analysed the classical and quantum relevance of the so-called \textit{fiducial cell} in homogeneous and isotropic cosmology, that is the compact region $V_o \subset \Sigma$ to which otherwise-divergent integrals over the non-compact spatial slice $\Sigma$ are restricted.~The assigned coordinate volume of the latter is $V_o$ and, in the limit $V_o \rightarrow \infty$, the region $V_o$ approaches again the spatial slice $\Sigma$.~While the volume $V_o$ is coordinate dependent, the quantity $\text{vol}(V_o) = \int_{V_o}\dd^3x\, a^3 = V_o a^3$ is not.~Such a fiducial cell is commonly assumed to be a non-physical auxiliary construction to be removed later on.~A closer inspection initiated in the present paper and complemented by a companion paper \cite{MeleOntheRoleof} where also a systematic study of the reduction to the homogeneous theory is presented, suggests that the fiducial cell is not so fiducial after all.~In agreement with some previous investigations \cite{BojowaldCanonicalderivationof,BojowaldMinisuperspacemodelsas,BojowaldTheBKLscenario,BojowaldEffectiveFieldTheory,BojowaldCriticalEvaluationof} based on a different starting point motivated by analogy with effective QFT,
$V_o$ has a physical interpretation as the scale on which homogeneity is imposed or in other words the scale on which periodic boundary conditions are imposed.~However, a physical length scale to this region can only be assigned after solving for the dynamical metric.

In the first part of the paper the classical theory was reviewed with emphasis on the way in which $V_o$ enters its canonical structure.~It was shown that it is possible to construct observables independent of the fiducial cell.~In particular, their dynamical evolution is also independent of it so that the on-shell physical predictions for classical dynamics are not affected by the choice of the fiducial cell.~In other words, changing the size of the homogeneity region (fiducial cell) is a symmetry of classical dynamics.~This is plausible as the regulator and compactification of the spatial slice are only relevant off-shell to construct the action, Lagrangian, Hamiltonian, and Poisson brackets, but not on-shell at the level of the Einstein equations.~As the local Einstein equations are sufficient to describe the full classical physics, it is not surprising that the size of $V_o$ does not affect the dynamics of these observables.~Only the off-shell quantities such as the Hamiltonian and the action diverge in the limit $V_o \rightarrow \infty$ where the regulator is removed.~Furthermore, studying the $V_o$-dependence of the canonical structure, it was argued that classically each fixed value of $V_o$ describes a different canonical theory and the different theories cannot be related by a canonical transformation.~Therefore, the regularisation leads to a whole family of classical homogeneous and isotropic theories, all leading to the same classical dynamics.

The fact that $V_o$ labels canonically inequivalent classical theories has important consequences at the quantum level.~In the second part of the paper, the quantisation of these classical theories was then studied in the framework of homogeneous and isotropic loop quantum cosmology.~It was argued that each value of $V_o$ corresponds to a different quantum representation characterised by different Hilbert spaces labelled by the $V_o$-values.~By relating states and expectation values within the Hilbert spaces carrying the different $V_o$-representations, it was possible to study the dependence on the change of $V_o$ of quantum operators which may also exhibit different scaling behaviours from their classical counterparts.~Moreover, it was possible to find an isomorphism between states of the Hilbert spaces associated to different $V_o$-regularisations, which preserves the action of the Hamiltonian and thus dynamics of quantum observables.~As discussed, this is not a necessity, but leads to a situation similar to the classical theory where dynamics is not dependent on the particular choice of fiducial cell and its coordinate volume.~In particular, as we have already noticed, the isomorphism proposed in this work comes to be the mapping sought in previous literature \cite{CorichiOnTheSemiclassical}.

The fact that the states so related have the same dynamics however does not mean that a change of $V_o$ has no effect in the quantum theory where not only dynamics, but also expectation values, higher moments, and quantum fluctuations are relevant.~In fact, two states $\psi^{(1)}$ and $\psi^{(2)}$ related as in \eqref{eq:isomorphygrav} can be physically distinct.~For example, the scaling behaviours \eqref{eq:scalingmomentsGR} of the elementary operators suggest that the point around which semiclassical states are peaked and their widths will transform under a change of $V_o$ \cite{MeleOntheRoleof}.~This might have consequences for the effective equations in that the classical trajectory over which the state is peaked could be rather different in the two quantum theories depending on the ratio $V_o^{(2)}/V_o^{(1)}$.

A quantum theory, moreover, also includes predictions about quantum fluctuations.~As we discussed with the help of the proposed mapping between the different quantum theories, these have a well-defined scaling behaviour when the $V_o$-representation is changed, both for the smeared and non-smeared relevant operators.~It was shown that the volume of $V_o \subset \Sigma$ in the quantum theory is actually independent of its coordinate value, which is consistent with diffeomorphism invariance and the fact that there is no reference field w.r.t.~which the volume of $V_o$ can be measured.~It was further argued that the quantum analogue of removing the regulator consists in choosing a state for which $\left<\widehat{\text{vol}\left(V_o\right)}\right>_\psi \rightarrow \infty$ and thus the physical volume assigned to the fiducial cell $V_o$ becomes infinite.~This is the case in late-time cosmology where the universe can be well approximated by the homogeneous model over large scales.~Nonetheless, by looking at observables respectively smeared over the fiducial cell $V_o$ and a subregion of it $V\subset V_o$, it was shown that, unlike the former, the quantum fluctuations of the latter explicitly depend on the size of the fiducial cell through the ratio $V/V_o$ interpreted as the (inverse) number of subcells $V$ homogeneously patched together into $V_o$.~In particular, the uncertainty relations for smeared \textit{physical observables} such as $\widehat{\text{vol}(V)}$ are affected as these would vanish in such a limit, leading to a seemingly classical theory even for a finite region $V$ as long as homogeneity can be safely imposed over large scales.~This observation is consistent and generalises the arguments in \cite{RovelliWhyAreThe}, where no distinction between $V$ and $V_o$ was considered.~In particular, this is in agreement with the observations that for large fiducial volumes the effective equations are valid \cite{TaverasCorrectionstothe,RovelliWhyAreThe,CorichiCoherentSemiclassicalStates,SinghSemi-classicalStates,FraserTamingFluctuationsfor}.~However, quantum fluctuations become relevant for a finite region, as e.g. in the early time regime, so that the fiducial cell is not playing the role of a mere regularisation but comes to be of physical relevance at the quantum level.

A possible interpretation of this is the fact that a classical theory is purely local, while a quantum theory is not.~The latter includes the description of non-local fluctuations and correlations.~The amount of fluctuations of a finite region $V \subset V_o$ are always measured w.r.t.~the fluctuations of the full fiducial cell $V_o$.~From this point of view, it is not surprising that the fluctuations for $V$ become negligible when the volume of $V_o$ becomes sufficiently large.~It should nevertheless be emphasised that $V$ can have still a finite volume as e.g.~the size of our universe, even if $\left<\text{vol}(V_o)\right>_\psi \rightarrow \infty$.~Thus, classically the fiducial cell has the \textit{physical} interpretation as the scale on which homogeneity and periodic boundary conditions are imposed.~Classical dynamics is independent of it, at least in the large volume regime where inhomogeneities can be safely neglected and the homogeneous description provides us with a good approximation.~The situation changes at the quantum level, where the scale of homogeneity has a physical effect on the quantum fluctuations.~Measuring the quantum fluctuations of $\widehat{\text{vol}(V)}$, allows to measure in principle the size of $V_o$, which is then a physical observable due to quantum effects.~Therefore, one needs to think carefully about the relevant quantum state which is studied and the volume that is assigned to $V_o$ in this way.~In the follow up paper \cite{MeleOntheRoleof}, we provide a more detailed analysis to understand these features form a full theory point of view by imposing homogeneity constraints systematically for the modes of classical fields, determining the resulting canonical structure of the symmetry-reduced classical theory, and studying its quantisation.

As a final note, we remark that changing $V_o$ at the quantum level also affects the relevant couplings, here the polymerisation scale $\lambda$, and forces to perform a very precise identification of states in order to be consistent with the dynamics.~This looks very much like a very na\"ive renormalisation procedure and in future work it would be interesting to relate this to previous work in the context of coarse graining and renormalisation in loop quantum cosmology \cite{BodendorferCoarseGrainingAs,BodendorferRenormalisationwithsu11,BodendorferPathintegralrenormalisation,HanImprovedmubarScheme}.~More specifically, as discussed in details in the follow up paper \cite{MeleOntheRoleof}, the classical homogeneous minisuperspace theory results from a twofold procedure consisting in neglecting the inhomogeneous modes with wavelength larger than the cell size and also in truncating those remaining inhomogeneous modes inside the cell so that one is left only with the zero mode.~When considering then multiple cells patched together into a bigger cell, the fully homogeneous analysis discussed here demands the zero modes in different cells to be equal while setting to zero all inhomogeneous modes.~This essentially amounts to replicate the physics in one cell into all the others and in turn leads to the mapping of states \eqref{eq:isomorphygrav} between different Hilbert spaces and no renormalisation of the Hamiltonian yet.~The present setting can be therefore extended in the following ways.~First, even neglecting interactions between the cells which in our framework should be encoded in boundary terms at the surfaces of the individual cells, the inclusion of inhomogeneous modes with wavelength larger than a single cell size would require us to impose different gluing conditions for the field modes.~Therefore, we expect the dynamics for states in the quantum theories of a single cell or many cells not to be the same any-more and, consequently, the mapping between different $V_o$-valued Hilbert spaces to be modified.~Second, a more complete picture would then require the interactions between neighbouring cells to be also included. In this respect, it would be interesting to compare the resulting analysis with previous work on perturbations around homogeneous cosmological spacetimes as initiated in \cite{Wilson-EwingLatticeLoopQuantum}, and systematically study the regime of validity of perturbative treatments of inhomogeneities at the quantum level.~We believe that the framework presented in this paper (and in its follow up paper \cite{MeleOntheRoleof}) offers interesting starting points for investigating these questions which are thus left for future investigation.

\section*{Acknowledgements}
The authors would like to thank Norbert Bodendorfer, Martin Bojowald, and Edward Wilson-Ewing for discussions and useful feedback on an early version of the paper.~The authors are also grateful to the anonymous referee whose questions helped improve the presentation.~JM would like to further thank Alejandro Perez for fruitful discussions.~The work of JM was made possible through the support of the ID\# 61466 grant from the John Templeton Foundation, as part of the First Phase \href{https://www.templeton.org/grant/the-quantum-information-structure-of-spacetime-qiss}{\textit{The Quantum Information Structure of Spacetime}} (QISS) Project.~The work of FMM was supported by funding from Okinawa Institute of Science and Technology Graduate University.~This project/publication was also made possible through the support of the ID\# 62312 grant from the John Templeton Foundation, as part of the Second Phase of \href{https://www.templeton.org/grant/the-quantum-information-structure-of-spacetime-qiss-second-phase}{\textit{The Quantum Information Structure of Spacetime}} Project.~The opinions expressed in this publication are those of the author(s) and do not necessarily reflect the views of the John Templeton Foundation.

%\bibliographystyle{utphysmendeley}
%\bibliography{library}

\end{document}